# Frequency-Dependent Photothermal Measurement of Transverse Thermal Diffusivity of Organic Semiconductors

J.W. Brill,[1] Maryam Shahi,[1] Marcia M. Payne,[2] Jesper Edberg,[3] Y. Yao,[1] Xavier Crispin,[3] and J.E. Anthony[2]

[1]Department of Physics and Astronomy, University of Kentucky, Lexington, KY 40506-0055, USA
[2]Department of Chemistry, University of Kentucky, Lexington, KY 40506-0055, USA
[3]Linköping University, Department of Science and Technology, Organic Electronics, SE-601 74 Norrköping, Sweden

We have used a photothermal technique, in which chopped light heats the front surface of a small ( ~ 1 mm$^2$) sample and the chopping frequency dependence of thermal radiation from the back surface is measured with a liquid nitrogen cooled infrared detector. In our system, the sample is placed directly in front of the detector within its dewar. Because the detector is also sensitive to some of the incident light which leaks around or through the sample, measurements are made for the detector signal that is in quadrature with the chopped light. Results are presented for layered crystals of semiconducting 6,13-bis(triisopropylsilylethynyl) pentacene (TIPS-pn) and for papers of cellulose nanofibrils coated with semiconducting poly(3,4-ethylene-dioxythiophene):poly(styrene-sulfonate) (NFC-PEDOT). For NFC-PEDOT, we have found that the transverse diffusivity, smaller than the in-plane value, varies inversely with thickness, suggesting that texturing of the papers varies with thickness. For TIPS-pn, we have found that the interlayer diffusivity is an order of magnitude larger than the in-plane value, consistent with previous estimates, suggesting that low-frequency optical phonons, presumably associated with librations in the TIPS side-groups, carry most of the heat.

## I. Introduction

The small molecule organic semiconductor 6,13-bis(triisopropylsilylethynyl) pentacene (TIPS-pn)[1] has received a lot of attention because of the ease with which it can be cast from solution into self-assembled films, e.g. for use in thin-film transistors. In the crystal, the pentacene backbones form a brickwork pattern in the **ab**-plane, with the TIPS side-groups projecting along the interlayer, **c**-axis direction.[1,2] Well-ordered films, with **c** approximately normal to the substrate, can be prepared by a number of techniques, including solution casting,[3] dip-coating,[4] ink-jet printing,[5] and solution shearing,[6] for which films with charge carrier (hole) mobilities > 10 cm$^2$/V·s were obtained.

For electronic applications, especially in submicron-size components, it is important that the thermal conductivity be sufficiently high (e.g. > $\kappa_0$ ~ 10 mW/cm·K) to minimize Joule heating and device degradation.[7] Because even crystalline organic crystals typically have thermal conductivities less than $\kappa_0$, it was not obvious how well TIPS-pn transistors would perform as their size was reduced. For example, rubrene, another high electronic mobility semiconductor[8] with a layered structure,[9] was found to have in-plane and interlayer thermal conductivities of 4 and 0.7 mW/cm·K, respectively,[10,11] values similar to semiconducting polymers.[12-14]

However, we recently reported the results of measurements of the in-plane and c-axis thermal conductivities, measured with ac-calorimetry and found values of $\kappa > \kappa_0$ for both directions.[15] For these measurements, the "front surface" of the sample was heated with chopped light and the temperature oscillations on the back surface measured with a thermocouple glued to the surface.[16] For values of $\kappa_{\text{in-plane}}$, measurements were made at a low chopping frequency while part of the front surface was blocked by a movable screen,[17] and we found, $\kappa_{\text{in-plane}} \approx 16$ mW/cm·K, close to the value of quasi-one dimensional organic metals with strong chain-axis $\pi$-bonding, such as TTF-TCNQ.[18] For the interlayer value, for which measurements depend on the chopping frequency dependence of the signal, we could only determine a lower limit, $\kappa_c >$ ~ 225 mW/cm·K, because the frequency response was limited by the response time of the thermometer.[15] This very large value of $\kappa_c$ and unusual anisotropy ($\kappa_c > \kappa_{\text{in-plane}}$) suggest that much of the heat is carried by low-frequency vibrations (presumably librations) in the isopropyl side groups which project between the planes.[1,2,15] For these optical phonons to have sufficient velocity to propagate, there must be relatively large interactions between the isopropyl groups on neighboring planes. This suggestion is in contrast to the usual assumption that the interlayer interactions can be treated as van-der-Waals bonds between essentially "rigid molecules," in which case only acoustic phonons are expected to have sufficient dispersion to carry heat between the layers.[15] Indeed, a calculation of the phonon thermal conductivity of pentacene, lacking the side groups, has shown that most of the heat is carried by acoustic modes.[19]

To put these large interlayer thermal conductivities on somewhat firmer experimental ground, we have remeasured $\kappa_c$ for TIPS-pn using a frequency-dependent photothermal (i.e. "ac-calorimetric"[11,16]) technique[20,21] which we have adapted for small (area < few mm$^2$) crystals. To test the technique, we also measured thin samples of a "paper" of nanofibrillated cellulose fibers coated with poly(3,4-ethylene-dioxythiophene):poly(styrene-sulfonate) (NFC-PEDOT).[14] The large ionic as well as electronic conductivity of this polymer blend have made it a promising material for supercapacitor and electrochemical sensor applications.[14]

In Section II of this paper, we describe our experimental technique in detail. In Sections III and IV, we describe the NFC-PEDOT and TIPS-pn samples and the experimental results on each.

## II. Experimental Technique

TIPS-pn crystals typically grow as needles, up to 1 cm long and 1 mm wide, but less than 100 μm thick in the interlayer direction. Crystals up to ~ 1 mm thick can be grown as described in Section IV, but these often have steps on their surfaces and also typically have areas < 5 mm$^2$. Conventional thermal conductivity techniques cannot be used for samples of these small dimensions, so we have used ac-calorimetry, but instead of measuring temperature oscillations with a thermometer glued to the sample,[11,16] we measure the oscillating thermal radiation from the sample surface.

Such photothermal measurements are typically done with a liquid-nitrogen cooled mercury cadmium telluride infrared detector. In conventional arrangements, the sample, with area ~ 1 cm$^2$, is mounted outside the detector dewar, and oscillating thermal radiation from the sample focused on the detector with a parabolic mirror. The sample is typically heated with chopped light from a laser with light out of the detector's spectral range.[20,21]

Because of the small size of TIPS-pn samples, we instead chose to mount the sample inside the detector dewar. It is glued, with thermally insulating glue, to a small aperture (1 – 5 mm$^2$), held at room temperature, ~ 1 cm in front of the wideband (0.9 – 22 μm) MCT detector. The front surface of the sample is illuminated with chopped light and the oscillating thermal radiation from the back surface measured by the detector, as shown in the schematic in Figure 1a. The results are normalized to the frequency dependence (magnitude and phase shift) of the detector response (only significant for frequencies below 20 Hz), measured by directly illuminating the detector with low intensity chopped light.

The frequency dependence of the oscillating temperature depends on the external ($\tau_1$) and internal ($\tau_2$) thermal time constants of the sample. $\tau_1 = C/K$, where C is the sample heat capacity and K is the heat conductance out of the sample[16] (i.e. by radiation and through the glue), and for all our samples $\tau_1 > 1$ second. Our measurements determine $\tau_2$, the time for heat to propagate through the sample:

$$\tau_2 = d^2/90^{\frac{1}{2}} \, D_{trans}, \quad (1)$$

where d = the sample thickness and the transverse thermal diffusivity $D_{trans} \equiv \kappa_{trans}/c\rho$, where c is the specific heat and ρ the density.[16] If the chopping frequency $\omega = 2\pi F \gg 1/\tau_1$, the complex oscillating temperature on the back of the sample is:[16,20]

$$T_{ac}(F) = 4P_{in}\chi/\{\pi C\omega[\sinh\chi \cos\chi \, (1-i) - \cosh\chi \sin\chi \, (1+i)]\}, \quad (2a)$$
$$\text{with } \chi \equiv (90^{\frac{1}{2}} \omega \, \tau_2/2)^{1/2} = d(\omega/2D_{trans})^{1/2}. \quad (2b)$$

Here $P_{in}$ is the power of the absorbed incident light and the phase of the oscillations are measured with respect to that of the incident chopped light. There are three approximations in deriving Eqtn. (2): i) the absorption length for the incident light is much less than the sample thickness, so that essentially all the heating is at the front surface of the sample;[20] ii) the average absorption length for thermal radiation is also much less than d, so that all the thermal radiation hitting the detector is from the back surface of the sample;[20] iii) heat flow through the sample is "one dimensional", i.e. the sample is heated uniformly and d << the lateral dimensions of the sample[16] (although larger values of d will mostly decrease $\tau_1$ as heat can escape the sides of the sample). We have spectroscopically checked that the first two approximations hold for both our TIPS-pn and NFC-PEDOT:PSS samples. Our light source is a 200 W quartz-halogen bulb. The light is fed through a flexible light pipe to a glass window on the detector dewar and then through a 3 mm diameter silvered glass tube to the sample to provide approximately uniform heating of the sample. We estimate the incident power on the sample to be ~ 10 mW, which leads to < 10 K dc heating of the sample.

Although most of the infrared light from the source is strongly attenuated in the glass lenses and windows, there is still a significant amount of incident radiation within the MCT detection range. Some of this leaks around (e.g. through the glue) or through the sample and is detected. We therefore only fit the detector response ($V_{ac}$, measured with a lock-in amplifier) that is in quadrature with the incident light:

$$F\, V_{ac}(F) \sin(\theta + \theta_0) = R\, \chi(\sinh\chi\, \cos\chi + \cosh\chi\, \sin\chi)/(\sinh^2\chi\, \cos^2\chi + \cosh^2\chi\, \sin^2\chi). \quad (3)$$

Here $\theta$ is the measured phase of the signal with respect to the leaked-light signal, whose phase is determined as described below. Fitting parameters are $\theta_0$, the error in setting the leaked light signal phase (usually a few degrees), amplitude R, and $\tau_2$, from which we determine the transverse diffusivity by Eqtn. (1).

The quadrature signal, Eqtn. (3), is plotted in Figure 1b as a function of frequency. At low frequencies, $T_{ac} \propto 1/F$ and $T_{ac}$ goes to zero (and oscillates) at high frequency ($\omega \gg 1/\tau_2$). Therefore (and to also assure that the mechanically chopped beam is close to a square wave), we found it convenient to limit our measurements to $F \leq 400$ Hz. If $1/2\pi\tau_2$ is sufficiently below 400 Hz, the detector signal at 400 Hz will be mostly due to the leaked light, and we set the lock-in amplifier phase here. In addition, the MCT detector has increased (1/F) noise at frequencies below ~ 50 Hz, so we tried to choose samples with 50 Hz $< 1/2\pi\tau_2 <$ 300 Hz.

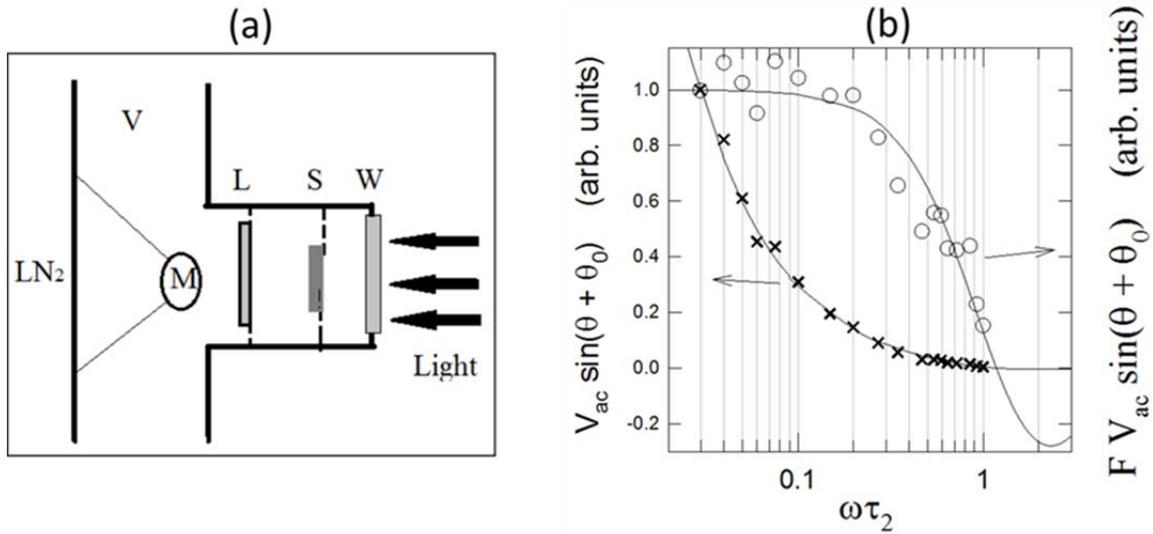

Fig. 1. a) Schematic (not to scale) of the apparatus. [V: liquid nitrogen dewar vacuum; W: glass window; S: sample; L: 10 μm longwave pass filter, used for the TIPS-pn and HOPG samples; M: MCT detector.] b) The solid curves show the theoretical frequency dependence of the thermal radiation in quadrature with the chopped light source as a function of frequency (Eqtn. (3)). Note that this quadrature signal changes sign near $\omega = 1/\tau_2$, i.e. qd ≈ 2, where $q \equiv (\omega/2D)^{1/2}$ is the thermal wavevector (Eqtn. (2b)). The symbols show the results for a d=230 μm thick sample of HOPG, with $\tau_2 = 1.2$ ms.

TIPS-pn crystals are relatively transparent in the near infrared, so the leaked light signal saturated the lock-in amplifier. (However, we estimate the *average* absorption length for the incident radiation to be ≈ 8 μm, much less than the thickness of the samples, so our approximation that most of the heating is at the front surface is valid.) For these samples, we placed a 10 μm longwave-pass filter between the sample to block most of the leaked light while passing ~ half of the thermal radiation from the sample. (The filter attenuated the leaked light signal by a factor greater than $10^5$. The filter's own thermal time constant ≈ 0.55 ms, similar to that of some samples, but its thermal signal is a few orders of magnitude smaller than that of the samples.) The NFC-PEDOT samples were much more opaque and the filter was not needed; for these, the ratio of the leaked light signal to the thermal signal at $\omega=1/\tau_2$ varied from ~ 1 to 20.

As an additional test of the technique, we measured the response of a sample [area ≈ 5 mm$^2$ and thickness = (230 ± 10) μm] of highly-oriented pyrolytic graphite (HOPG), for which the interlayer (c-axis) thermal conductivity has been reported to be ≈ 80 mW/cm·K, more than 200 times smaller than the in-plane value.[22] Because of the high reflectivity of HOPG, the signal, shown in Figure 1b, is relatively small and noisy, but our fit to Eqtn. (3) gave $\tau_2 = 1.2 \pm 0.2$ ms, corresponding to $\kappa_c = (77 \pm 20)$ mW/cm·K.

### III. NFC-PEDOT

Poly(3,4-ethylene-dioxythiophene):poly(styrene-sulfonate) (PEDOT:PSS) is a polymer blend that was developed for its electronic and thermoelectric properties.[23] Blending dimethylsulfoxide (DMSO) enhances phase separation of excess PSS and consequently improves "π-stacking" of PEDOT strands and the electronic conductivity by "secondary doping",[13,14] but these films are not mechanically strong. For our samples, PEDOT:PSS was blended with DMSO, glycerol, to improve plasticity and hygroscopicity (and ionic conductivity), and nanofibrillated cellulose (NFC) to form mechanically strong "papers". The cellulose acts as a scaffolding for the PEDOT:PSS, which clad the ~ 10 nm diameter, ~ 2 μm long NFC nanofibrils. The DMSO and glycerol molecules are dispersed between the disordered, entangled fibrils.[14]

We previously reported on values of the density ($\rho = 1.26$ g/cm$^3$), specific heat (c = 1.3 J/g·K) and in-plane diffusivity ($D_{\text{in-plane}} = 0.7$ mm$^2$/s).[14] The values of the specific heat and in-plane thermal conductivity are very similar to that of blends of cellulose with a non-conducting polymer.[24] However, it was expected that the transverse diffusivity would be smaller, because the cellulose fibrils tend to lie in the plane of the paper.[14]

While large area (> 1cm$^2$) samples of NFC-PEDOT:PSS samples are available, we cut much smaller samples (~ 1 mm$^2$) to test our photothermal technique. The material has several advantages for our photothermal measurement: a) samples of several thicknesses (measured to ± 1 μm) were available; b) the low diffusivity meant that small thickness samples could be used to keep $1/2\pi\tau_2 < 300$ Hz, decreasing the sample heat capacity and increasing the value of $T_{ac}$ (see Eqtn. 2a); c) as mentioned above, the samples are extremely opaque, so it was not necessary to use the longwave-pass filter.

Results for six samples of three different thicknesses, along with the fits to Eqtn. (3), are shown in Figure 2. The results for the two samples of each thickness are very reproducible, but the fitted values of $\tau_2$ had a much stronger dependence on thickness than the expected quadratic dependence (Eqtn. (1)). The resulting thickness dependence of the transverse diffusivity is shown in the Figure 2 inset. Note that since the typical nanofibril length (~ 2 μm) is much less than the sample thicknesses, $D_{trans}$ is expected to be independent of d if samples of different thickness have identical structures and compositions. Therefore, the observed inverse relation between diffusivity and thickness that we observe is probably due to thickness dependent water content and/or degree of nanofibril alignment. In particular, our results suggest that the nanofibrils are more disordered (i.e. less aligned in the plane) for thinner samples than thicker ones. (Investigations of the detailed structure of these papers are under consideration.) Note that for even the thinnest sample, $D_{trans} < D_{in-plane} = 0.7$ mm$^2$/s.

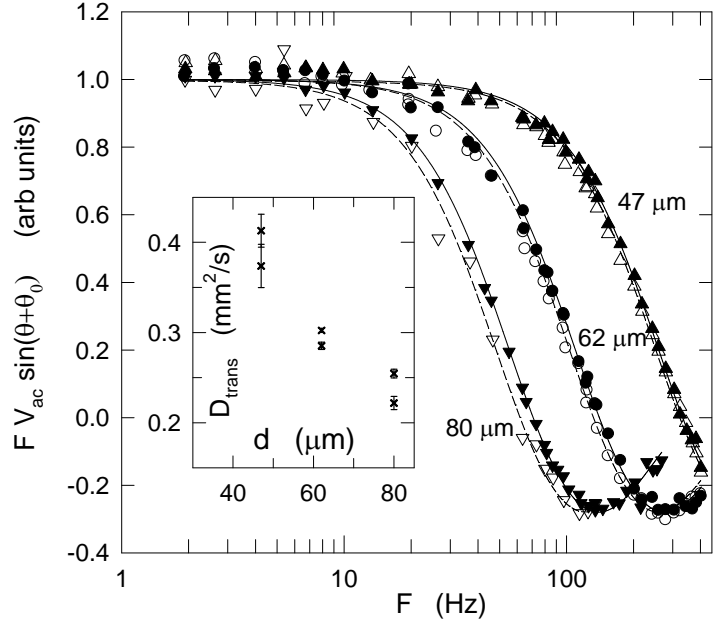

Fig 2. Measured values of $F V_{ac} \sin(\theta+\theta_0)$ and fits to Eqtn. (3) for six NFC-PEDOT:PSS samples of three thicknesses, as indicated. For each thickness, one sample is shown with solid symbols and its fit with a solid curve and the second sample with open symbols and a dashed curve. Inset: Values of the transverse diffusivity vs. thickness for the six samples; the error bars are due to the uncertainties in the fitted values of $\tau_2$.

## IV. TIPS-pn

Because of its large value of interlayer (i.e. transverse, **c**-axis) thermal conductivity, it was necessary to work on TIPS-pn crystals between 200 and 600 μm in thickness (e.g. to keep $1/2\pi\tau_2$ < 300 Hz). TIPS-pn was prepared as described in the literature,[25] and initially purified by recrystallization (three times) from acetone. To grow large, thick crystals for the studies reported here, the recrystallized TIPS-pn was added to boiling 2-butanone, butyl acetate, or 1-chlorobutane until the solution was saturated. The solution was then filtered quickly while hot through a fine glass frit, and the filtrate re-heated to reflux. The solution was then capped and placed in a dark, vibration-free environment to cool slowly. The crystals were allowed to grow in this environment for a period of 6 - 17 days. At the end of that time, the remaining solvent was decanted and the crystals harvested from the bottom of the growth container.

The thick crystals generally had irregular surfaces, i.e. not constant thicknesses; results are shown here for four samples with fairly well-defined values of d, but the uncertainty in thickness was the main source of uncertainty in determining the diffusivity, as shown below in Figure 4. (In addition, many crystals were "hollow"; these were eliminated by comparing the measured thickness from that determined from the mass.) Because of the larger thicknesses as compared to NFC-PEDOT samples, the heat capacities were much larger and therefore the signals smaller (see Eqtn. 2a); signals were further reduced (by ~ 50%) because of the need to use the longwave-pass filter. The 1/f detector noise therefore becomes very apparent (see Figure 3) for the thicker crystals; the data shown is the result of averaging several data sets to reduce noise.

Results for the four samples, together with their fits, are shown in Figure 3, and the fitted values of $\tau_2$ are shown in Figure 4. The linear variation of $\tau_2$ with $d^2$ shows that finite thickness effects (i.e. non-one-dimensional heat flow) are not significant for these samples, and from Eqtn. (1) we find the interlayer diffusivity: $D_c = (13 \pm 6)$ mm$^2$/s; using[1,15] $c = 1.48$ J/g·K and $\rho = 1.1$ g/cm$^3$, this corresponds to $\kappa_c = (210 \pm 100)$ mW/cm·K, similar to the lower limit concluded in Reference 15.[25]

This large interlayer thermal conductivity has not been previously observed in a "molecular crystal", for which the interlayer bonding is generally considered to be due to van-der-Waals interactions between essentially rigid molecules. It is not observed in pentacene crystals,[19,26] for which there are no side-groups projecting between the planes, nor in rubrene,[10,11] for which tetracene backbones align in the plane with relatively rigid phenyl side groups projecting between the planes.[9] This suggests that the large interlayer thermal conductivity of TIPS-pn is associated with the ability of the floppy TIPS side-groups to conduct heat. This is supported by the fact that we observed similarly high values of $D_c$ for other materials with the same or similar interlayer side-groups.[15]

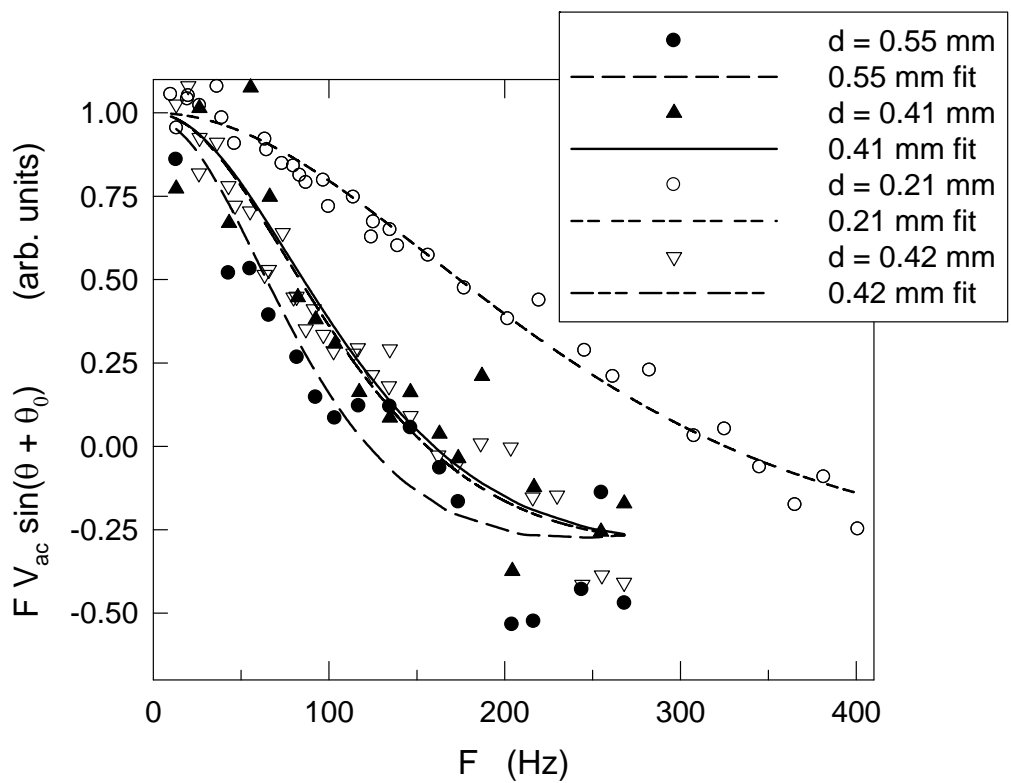

Fig 3. Measured values of F $V_{ac} \sin(\theta+\theta_0)$ and fits to Eqtn. (3) for four crystals of TIPS-pn, with thicknesses, as indicated. The scatter in the data for each sample shows the uncertainty of the measurements.

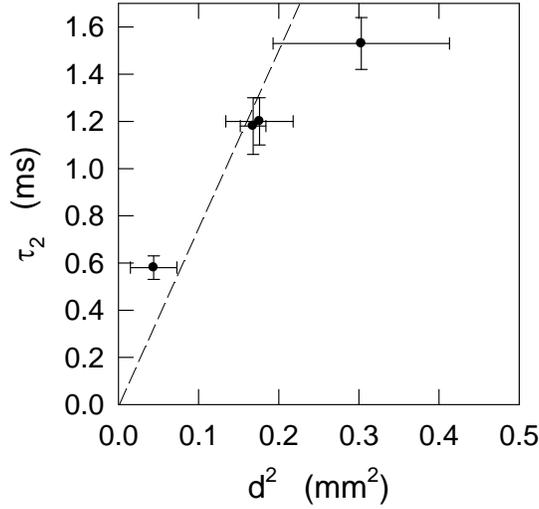

Fig. 4. Fitted values of $\tau_2$ vs. thickness squared for the four TIPS-pn crystals of Figure 3. The uncertainties in $d^2$ reflect the non-uniform thicknesses of the samples while the uncertainties in $\tau_2$ show the resulting uncertainties in the fits. The dashed line shows the expected quadratic dependence; $D_c = 1/(\sqrt{90} \cdot \text{slope})$.

In fact, kinetic theory considerations show that treating the molecules as rigid and considering only acoustic phonon propagation as a mechanism for heat conduction is extremely unlikely to account for the high thermal conductivity. In kinetic theory, the phonon thermal conductivity is expressed in terms of the sum over phonon modes (j) of the product of the specific heat, velocity, and mean-free path for each mode:

$$\kappa = (\rho/3) \, \Sigma \, c_j \, v_j \, \lambda_j \quad (4).$$

If it is assumed that only acoustic modes have sufficient velocity to make a contribution to the thermal conductivity, then $\kappa = (\rho/3) \, c_{acoustic} \, \langle v_{acoustic} \, \lambda_{acoustic} \rangle$.[11] At room temperature, $c_{acoustic} \approx 3R/M$, where R is the gas constant and M is the molecular weight (638 g/mole). Assuming a typical value of $v_{acoustic} \sim 2$ km/s gives $\lambda_{acoustic} \sim 700$ nm $\sim 400 \, d_c$, where $d_c = 1.7$ nm is the interlayer spacing. Such a large mean-free path is extremely unlikely in view of the measured large thermal disorder, e.g. shear motion of the molecules.[27]

However, since the librational modes typically have energies $\leq k_B T_{room}$,[28] they can also carry heat at room temperature if they have sufficient dispersion. Furthermore, because of the large number of terminal methyl groups (12 on each molecule), they can potentially carry an order of magnitude more heat than the acoustic modes alone. In fact, the quadrupolar coupling between these groups may give librational phonons sufficient velocity to contribute, assuming a typical quadrupole moment $Q \sim 10^{-39}$ C·m$^2$.[29] Since the distance between isopropyl groups on neighboring layers is r ~ 0. 4 nm, the interaction energy $U_{quad} \sim Q^2/(4\pi\varepsilon_0 r^5) \sim 5$ meV.[28] This bandwidth would give a librational optical phonon velocity close that of acoustic phonons: $v_{lib} \sim U_{quad}\, d_c/h \sim 2$ km/s, where $h$ = Planck's constant. Of course, direct proof of propagating low-energy optical phonons in TIPS-pentacene and related materials would require inelastic neutron or x-ray measurements of phonon dispersion, which would be difficult in the small, low-Z materials. Indirect proof may come from measurements of in-plane and interlayer thermal diffusivity in materials with a variety of interlayer constituents and structures.

It is also noteworthy that the phonon mobility of TIPS-pn, as measured by the thermal conductivity, has the opposite anisotropy from the electronic mobility. While the **c**-axis electrical conductivity has not been measured, a band structure calculation has indicated extremely flat-bands (bandwidths << 10 meV) in the interlayer direction but in-plane electron and hole bandwidths ~ 300 and 150 meV, respectively.[30]

In summary, we have used a modified frequency-dependent photothermal technique, in which the sample is placed directly in front of an MCT detector in the detector dewar, to measure the transverse thermal diffusivity of small samples. The simplified geometry of the technique allows samples with areas as small as 1 mm$^2$ to be measured. It is ideally suited for materials with small thermal conductivities, such as polymeric samples, and results are presented for NFC-PEDOT composites. However, we have also used it, with poorer signal/noise, for crystalline TIPS-pn, which has a very large interlayer thermal diffusivity. Its large value of $D_c$ shows that interactions between low-energy optical phonons can greatly increase the thermal conductivity of molecular crystals.

ACKNOWLEGEMENTS
We thank Abdellah Malti and Zia Ullah Kahn for discussions of the properties of NFC-PEDOT and Doug Strachan, Mathias Boland, and Mohsen Nasseri for providing the HOPG sample. This research was supported in part by the National Science Foundation, Grant No. DMR-1262261 (JWB), the Office of Naval Research, Grant No, N00014-11-0328 (JEA), and The Knut and Alice Wallenberg foundation (Power Paper project) KAW 2011.0050 (XC).